\documentclass[aip,jmp,amsmath,amssymb,preprint]{revtex4-1}
\usepackage{graphicx}
\usepackage{dcolumn}
\usepackage{bm}

\begin{document}


\title{Enhanced spin accumulation at room temperature in graphene spin valves with amorphous carbon interfacial layers}

\author{I. Neumann}
\affiliation{ICN2 - Institut Catala de Nanociencia i Nanotecnologia, Campus UAB, Bellaterra, 08193 Barcelona, Spain}
\affiliation{Universitat Auton\'oma de Barcelona, Bellaterra, 08193 Barcelona, Spain}
\author{M. V. Costache}
\affiliation{ICN2 - Institut Catala de Nanociencia i Nanotecnologia, Campus UAB, Bellaterra, 08193 Barcelona, Spain}
\author{G. Bridoux}
\affiliation{ICN2 - Institut Catala de Nanociencia i Nanotecnologia, Campus UAB, Bellaterra, 08193 Barcelona, Spain}
\author{J. F. Sierra}
\affiliation{ICN2 - Institut Catala de Nanociencia i Nanotecnologia, Campus UAB, Bellaterra, 08193 Barcelona, Spain}
\author{S. O. Valenzuela}
\affiliation{ICN2 - Institut Catala de Nanociencia i Nanotecnologia, Campus UAB, Bellaterra, 08193 Barcelona, Spain}
\affiliation{Universitat Auton\'oma de Barcelona, Bellaterra, 08193 Barcelona, Spain}
\affiliation{ICREA - Instituci\'o Catalana de Recerca i Estudis Avan\c{c}ats, 08010 Barcelona, Spain}



\begin{abstract}
    We demonstrate a large enhancement of the spin accumulation in monolayer graphene following electron-beam induced deposition of an amorphous carbon layer at the ferromagnet-graphene interface. The enhancement is $10^4$-fold when graphene is deposited onto poly(methyl metacrylate) (PMMA) and exposed with sufficient electron-beam dose to cross-link the PMMA, and $10^3$-fold when graphene is deposited directly onto SiO$_2$ and exposed with identical dose. We attribute the difference to a more efficient carbon deposition in the former case due to an increase in the presence of compounds containing carbon, which are released by the PMMA. The amorphous carbon interface can sustain very large current densities without degrading, which leads to very large spin accumulations exceeding 500 $\mu$eV at room temperature.
    \\

\end{abstract}


\maketitle


Graphene has attracted the attention of the spintronics community due to the long spin lifetimes and long spin relaxation lengths expected from its small intrinsic spin-orbit coupling and the lack of hyperfine interaction with the most abundant carbon nuclei ($^{12}$C).\cite{castroneto09, dery12} Nonlocal spin valves\cite{js85, SOV09} (NLSV) comprising ferromagnetic contacts and a graphene channel\cite{tombros07} are of particular interest because of the ease to manipulate the spin during transport by external electric fields or by modifying the graphene physical properties through the addition of adatoms.\cite{castroneto09,balakrishnan13} They can also be used to study spin torque switching\cite{kimura06,otani10} or spin Hall effects,\cite{maekawa12} if large spin accumulation and large pure spin currents are achieved.

Depending on the interface characteristics between the ferromagnet (FM) and graphene, graphene NLSVs have been classified into three types: those having Ohmic, pinhole or tunneling contacts.\cite{han10} Because of the so-called conductance mismatch and the spin absorption at both injector and detector FMs, the spin injection efficiency, i.e. the effective spin polarization, is strongly suppressed for Ohmic contacts. Typical reported nonlocal spin magnetoresistances in this case, i.e. the overall change $\Delta R_\mathrm{NL}$ in the nonlocal spin resistance between the parallel and antiparallel configuration of the electrodes magnetizations, are in the range of a few mOhms to a few tenths of mOhms.\cite{han09a} Larger $\Delta R_\mathrm{NL}$'s have been obtained by placing an insulator between graphene and the FMs, which helps circumvent the conductance mismatch, and reduce the spin absorption in the latter.\cite{tombros07,han10} The used insulators are typically MgO or AlO$_x$, because of their success for tunnel magnetoresistance.\cite{moodera95,miyazaki95,parkin04,yuasa04} In this way, $\Delta R_\mathrm{NL}$ was observed to increase to up to a few Ohms (pinhole barrier) or a hundred Ohms (tunnel barrier).\cite{han10} However, high-resistance tunnel barriers are detrimental for high-speed and spin-torque applications and alternative approaches to increase $\Delta R_\mathrm{NL}$ and the spin accumulation have been proposed both in metallic systems, for example by adding a native oxide layer at a Ni$_{80}$Fe$_{20}$ /Ag interface\cite{mihajlovic10} or by increasing confinement,\cite{laczkowski12} and in graphene, for example by adding a thin Cu interfacial layer at the metal-graphene interface.\cite{zhang12}

Here we investigate FM/aC/graphene junctions as a spin polarizer, where aC stands for amorphous carbon. The transport properties of metal/aC/graphene interfaces have not been studied, much less its spin transfer properties. However, previous demonstrations\cite{yoshikawa07} of improved metal-nanotube contacts using electron-beam induced deposition (EBID) of aC and the fact that carbon is a light material, which may introduce relatively low spin dephasing, make aC an excellent candidate for spintronic applications. Indeed, we demonstrate a $10^4$-fold enhancement of the spin signal in graphene lateral spin valves following EBID of aC interfacial layers. The interfaces are very robust, simple to fabricate, and can sustain very large currents without degradation, which allows us to generate spin accumulation with unprecedented magnitude ($> 500$ $\mu$V at room temperature).


We fabricated three specific types of devices of equal dimensions, in the following referred to as A, B and C. Type A devices, which we use as a reference, are graphene NLSVs with cobalt electrodes and Ohmic contacts, as reported in previous studies.\cite{han09a,han09b,shiraishi09} Here, graphene is directly exfoliated onto a $p^+$ Si/Si$\mathrm{O_2}$ substrate (440\,nm oxide thickness) and then suitable flakes are localized with an optical microscope. Raman spectroscopy is used to pre-calibrate the microscope image contrast in order to identify single-layer graphene flakes. The cobalt electrodes (26\,nm thick) are defined using electron-beam lithography; cobalt is deposited using an electron-beam evaporator with a base pressure of about 1$\times10^{-7}$\,Torr.

\begin{figure}[ht]
\includegraphics[width=9cm]{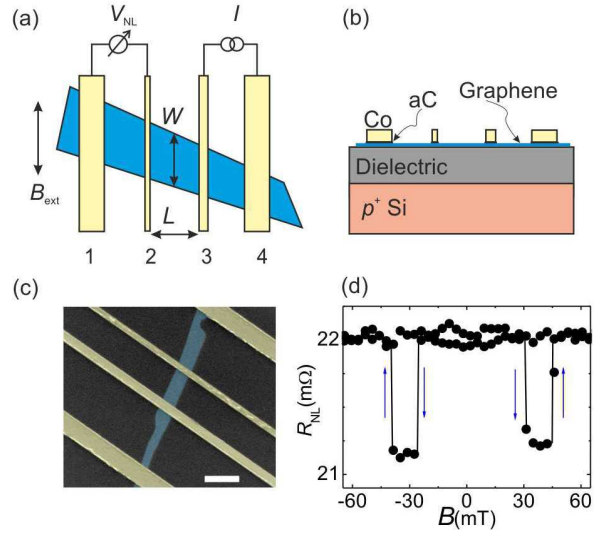}
\vspace{3mm}
\caption{\label{Fig_1}
Device layout, top (a) and lateral (b) views. Four ferromagnetic electrodes (yellow) are in contact with graphene (blue). The dielectric consists of 440\,nm of Si$\mathrm{O_2}$ for type A and B devices, and 285\,nm of Si$\mathrm{O_2}$ plus 200\,nm of PMMA for type C devices. A layer of amorphous carbon (aC) is deposited with EBID at the interface between graphene and Co for type B and C devices. (c) SEM image of a device, the scale bar equals 1$\,\mathrm{\mu m}$ (d) NLSV measurement for a reference device with transparent contact (type A).}
\end{figure}

For type B and C devices, we deposit an aC layer in the contact area just after exfoliation. Amorphous carbon is deposited by EBID, which consists in using a focused electron-beam (e-beam) that decomposes molecules, such as hydrocarbons, that are then adsorbed onto graphene. This process is well established\cite{vandorp08} and has been used, for example, to fabricate complex carbon structures,\cite{lemoine07} conducting bridges,\cite{rykaczewski10} and contacts with carbon nanotubes.\cite{yoshikawa07} Carbon precursors can either be introduced externally using a gas source or simply be present as residual hydrocarbons at the background pressure of a scanning electron microscope (SEM) chamber.\cite{vandorp08} Carbon deposition from residual hydrocarbons onto graphene was recently demonstrated by direct visualization in a transmission electron microscope.\cite{barreiro13}

We perform EBID in our e-beam lithography system. We use an e-beam area dose of 9000\,$\mu \mathrm{C}/\mathrm{cm^2}$ and an accelerating voltage of 30\,keV. This dose is about 15-20 times larger than that required to expose PMMA for e-beam lithography, and is large enough to deposit a thin layer of aC at the residual pressure of our system,\cite{vandorp08, lemoine07, rykaczewski10, yoshikawa07} which is in the $10^{-6}$\,Torr range. For type B devices, we use the same substrate as for type A. For type C devices, we introduce an additional 200\,nm thick PMMA layer between graphene on a $p^+$ Si/Si$\mathrm{O_2}$ substrate with 285\,nm Si$\mathrm{O_2}$. Here, the EBID dose cross-links the PMMA, making it resistant to acetone, which is used during the lift-off process after the Co contact deposition. PMMA is known to be a suitable high-$\kappa$ dielectric substrate for graphene devices,\cite{ponomarenko09} as well as for the fabrication of insulating or hydrophobic layers.\cite{barthold11} It also increases the presence of carbon-rich molecules in the environment during EBID, therefore changing the aC EBID dynamics.\cite{vandorp08}

The device design and the nonlocal spin valve measurement scheme \cite{js85, SOV09, IN13} are shown in Figs. 1(a) and 1(b). The distance $L$ between the inner contacts is kept constant at 1.15$\,\mathrm{\mu m}$ for all devices, while the width of the graphene $W$ varies between 500$\,\mathrm{nm}$ and 1$\,\mathrm{\mu m}$. The widths of the ferromagnetic electrodes determine their coercive fields. The inner electrodes, 2 and 3, are 100\,nm and 200\,nm wide, respectively, while the outer ones, 1 and 4, are both 500\,nm wide. A current $I$ is injected between two of the ferromagnetic electrodes (3 and 4) resulting in a nonlocal voltage $V_\mathrm{NL}$ over the detector electrodes (1 and 2). Application of an in-plane, external magnetic field $B$ along the axis of the ferromagnets allows us to switch their magnetizations sequentially. As we sweep $B$, a change in the nonlocal spin resistance $R_{\mathrm{NL}}= V_{\mathrm{NL}}/I$ occurs when the relative orientation of the magnetizations of the inner ferromagnets switches from parallel to anti-parallel, as in the NLSV measurement of a reference sample (type A) shown in Fig. 1 (d). Here, $\Delta R_{\mathrm{NL}}$ is about $1\,\mathrm{mOhm}$, in agreement with previously reported values for transparent contacts.\cite{han09a,han09b,shiraishi09} All measurements presented in this paper were carried out at room temperature.

\begin{figure}[ht]
\includegraphics[width=8.5cm]{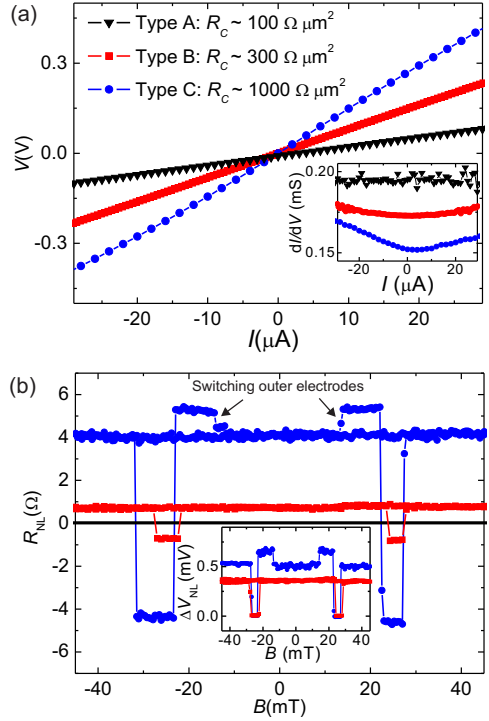}
\vspace{-3mm}
\caption{\label{Fig_2}
(a) Typical IV-curves for the three types of devices: type A (black triangles), type B (red squares) and type C (blue circles). Inset: Corresponding \textit{dI/dV}-curves, offset for clarity (+50\,$\mu$S for type B, +85\,$\mu$S for type C). (b) NLSV measurements for the devices in (a). $\Delta R_{\mathrm{NL}}$ for type C devices is roughly four orders of magnitude larger than for type A devices. Inset: NLSV measurements for $I=400\,\mu$A (type B) and $I=46\,\mu$A (type C). The data was displaced vertically to stress the overall change of $V_{\mathrm{NL}}$, $\Delta V_{\mathrm{NL}}$.}
\end{figure}


Figure 2 shows our main results. Amorphous carbon deposition by EBID leads to an increase in the contact resistance per unit area, $R_C$, between the ferromagnet and graphene [Fig. 2(a)] and dramatically enhances the nonlocal spin signal [Fig. 2(b)]. We performed IV-measurements in 2-point configuration between pairs of ferromagnetic electrodes [Fig. 2(a)]. Even though it is not straightforward to accurately determine the contact resistance between graphene and a metallic electrode,\cite{xia11} our measurements over more than 20 devices demonstrate that $R_C$ systematically increases from A to B to C devices. After subtracting the resistance from the leads and graphene (measured in four point configuration),\cite{blake09} we estimate that $R_C$ is $\leq 100\,\Omega \mu\mathrm{m}^2$ for type A devices, as observed previously.\cite{han09a} For type B and C devices, $R_C$ increases significantly to about $300$ and $1000\,\Omega \mu\mathrm{m}^2$, respectively. Numerical differentiation of the IV-measurements [inset of Fig. 2(a)] reveal nonlinearities in these devices that are not observed in the type A ones, which is an indication of differences in the character of the electronic transport. Previous studies in metal-carbon nanotube contacts fabricated by EBID presented similar features, which were associated to a combination of tunneling and ohmic resistances.\cite{yoshikawa07} However, as in the case for nanotubes, further studies are required to identify the precise nature of our FM/aC/graphene contacts.

The thickness of the amorphous carbon layer can be roughly estimated by assuming Ohmic behavior and using typical resistivity values of EBID-grown aC films, \cite{rykaczewski10} $\rho^{aC} \sim 2 \times 10^{5} \,\Omega \mu$m. Considering an increase of $\sim 100\,\Omega \mu\mathrm{m}^2$ and $800\,\Omega \mu\mathrm{m}^2$ in the contact resistance (after subtracting $ 100\,\Omega \mu\mathrm{m}^2$ per interface) we calculate that the aC thickness for type B and type C devices is about 0.5 nm and 4 nm, respectively. These values likely represent an upper limit for the thickness because roughness in the aC films and tunneling transport would effectively increase the contact resistance. It is also plausible that a small amount of carbon on graphene changes the deposition dynamics of the cobalt that follows, leading to a different structure at the interface and, perhaps, to different characteristic resistance and polarization.\cite{zutic13} The coexistence of two structures with similar energy was recently observed in graphene on Ni(111).\cite{zhao11}

NLSV measurements for typical A, B and C devices are shown in Fig. 2(b) in the same scale. We have found that $\Delta R_{\mathrm{NL}}$ for type B devices varies from hundreds of $\mathrm{mOhms}$ to the lower Ohms range, which is three orders of magnitude larger than the values for our type A devices. The enhancement is so large that the features of the measurements shown in Fig. 1 (d) cannot be resolved in Fig. 2 (b) and appear as a straight line. For type C devices, $\Delta R_{\mathrm{NL}}$ is even larger, typically about ten Ohms ($ \approx 8\,\mathrm{Ohms}$ for the device in Fig. 2). This represents an additional order of magnitude increase and, therefore, up to a $10^4$-fold overall enhancement when comparing with type A devices. $\Delta R_{\mathrm{NL}}$ in type B and C devices compares well with the reported values for pinhole contacts using MgO or AlO$_x$.\cite{tombros07, han10, yang11} Notably, very high-current densities can be applied to our contacts without deteriorating them. As shown in the inset of Fig. 2 (b), we are able to achieve very large absolute nonlocal spin voltages of about 500$\,\mu$V, which is the largest value reported to date in any material.\cite{SOV05,shiraishi09,otani10}

The introduction of disorder in graphene by the e-beam is unlikely at an acceleration voltage of 30\,keV, which is below the knock-on damage threshold of carbon nanostructures.\cite{barreiro13, meyer12} This agrees with the fact that we found no correlation between the carrier mobility of the graphene sheet and the exposure to the e-beam dose, even when graphene is fully exposed. Indeed, graphene on cross-linked PMMA frequently exhibits higher mobility and lower residual doping than graphene on Si$\mathrm{O_2}$. The mobilities of the above devices were of about 2000 to 3000 $\mathrm{cm^2/Vs}$ but in some cases it can exceed 20000 $\mathrm{cm^2/Vs}$ for fully exposed graphene.

We also performed spin precession (Hanle) measurements to determine the spin relaxation length $\lambda_{sf}$ of the type A and B devices in Fig. 2. Such measurements were not possible for type A devices because of the small signal and the large spin absorption by the contacts. By fitting the measurements to a one-dimensional model,\cite{js85,tombros07} we obtain $\lambda_{sf} \approx 1.3\,\mathrm{\mu m}$. The distance between the contacts is therefore smaller than $\lambda_{sf}$ and minor changes in $\lambda_{sf}$ cannot change the magnitude of $R_{\mathrm{NL}}$ significantly. The Hanle measurements also deliver the effective polarization $P$ of the electrodes and the spin lifetime $\tau_{sf}$ in graphene. For the devices in the present paper, $\tau_{sf}$ is smaller for device C (85 ps) than for device B (145 ps). However, $\tau_{sf}$ is in the range of 100 to 200 ps for most devices and we found no clear correlation between the spin lifetime and the type of contact (B or C) or type of substrate (PMMA or SiO$_2$). On the other hand, $P$ can be up to 10 to 15\% for both B and C devices, but tends to be smaller for the former, where it can be as low as a few percent. The extracted values of $P$, $\tau_{sf}$ and $\lambda_{sf}$ are of the same order to those observed in devices with similarly short injector/detector separation ($\sim 1 \mu$m) or with pinhole barriers, where contact dephasing might play a role.\cite{tombros07,maassen12}

We thus argue that the increase in $R_C$ is solely due to EBID. In the case of type B devices, the aC originates from the hydrocarbons present in the chamber of the e-beam lithography system, as previously observed.\cite{vandorp08, lemoine07, rykaczewski10, yoshikawa07} The additional increase in $R_C$ for type C devices might be associated to the release of carbon-rich molecules from the PMMA layer, which may act as precursors and decompose in the electron-beam irradiated area, resulting in a larger aC-deposition rate than at the residual chamber pressure.\cite{vandorp08}

Despite the fact that no signs of degradation of the graphene sheet are observed after EBID, if possible, one should perform the EBID step in the contact region only, as shown in Fig. 1 (b), which leaves the graphene between the contacts completely unaffected. This could be relevant for efficient cleaning of the graphene sheet because, as recently pointed out,\cite{barreiro13} amorphous carbon might leave residues even after current annealing following EBID.

\begin{figure}[ht]
\includegraphics[width=8cm]{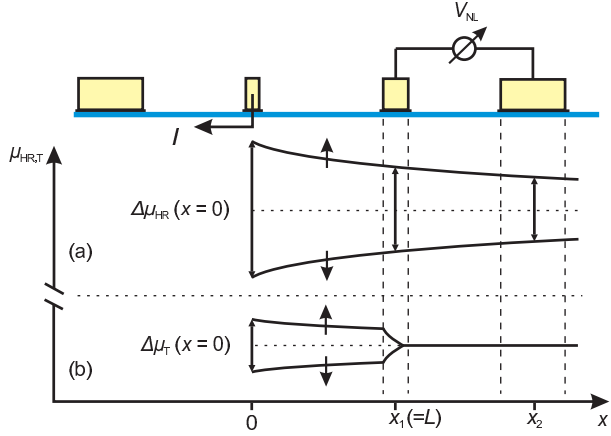}
\vspace{3mm}
\caption{\label{Fig_4}
Schematics of the spin-up and spin-down electrochemical potentials in graphene for highly resistive (a) and transparent (b) contacts.}
\end{figure}

Finally, an additional step in $R_{\mathrm{NL}}$ as a function of $B$, which is due to the switching of the outer electrodes, becomes apparent in the NLSV measurements of our type C devices [Fig. 2(b); $B\sim 15$\,mT]. This feature is well-known.\cite{marius06} For wide contacts, it is only present when a sufficiently large contact resistance prevents the spin-absorption effect. In Fig. 3, we qualitatively show the variation of the electrochemical potential $\mu_{HR,T} (x)$ for spin-up and spin-down electrons for transparent (T) and highly-resistive (HR) contacts at the detector electrodes, corresponding to type A and type C devices. If the contact resistance is high [Fig. 3(a)], no contact induced spin relaxation occurs and, therefore, when the magnetization of the outer detector switches, $V_{NL}$ changes by $\Delta V_{NL} \propto \Delta \mu_{HR} (x=x_2)$. In the case of transparent contacts [Fig. 3 (b)], the effect of the spin absorption by the ferromagnet is two-fold. The overall spin accumulation is smaller and, for wide enough contacts, it is completely suppressed below the contact. In this situation, the switching of the outer detector electrode does not affect the measurements because $\Delta V_{NL} \propto \Delta \mu_{T} (x=x_2) = 0$. An analogous argument can be made in relation to the second injector.
Thus, the fact that this feature occurs most notably for type C devices further corroborates our hypothesis of the formation of an aC interface layer that increases the contact resistance between Co and graphene and leaves graphene unaffected.

In conclusion, we have implemented graphene-based NLSVs. Nonlocal measurements show that an amorphous carbon layer at the FM/graphene interface, which is deposited by e-beam induced deposition, can result in a large enhancement in the spin injection/detection efficiency, even at large applied injection currents. We found a $10^4$-fold enhancement in comparison to ohmic contacts, but improvements can be expected after optimizing the deposition of carbon by choosing the appropriate carbon precursor and by controlling its quantity in a suitable electron beam lithography system. Nevertheless, further studies are needed to precisely determine the nature of the interface, which can have ohmic or tunneling character or a combination of both.

After finishing this work, and in order to test the transferability of our methods, we have repeated the amorphous carbon deposition procedure in a second electron-beam lithography system from a different vendor and found essentially the same results. This underscores the importance of amorphous carbon for future spintronic research, specially because of the simplicity and transferability of the deposition method and the low reactivity of carbon. Amorphous carbon can be used as an alternative material to conventional insulators used in spintronics, such as MgO or AlO$_x$. In particular, it might open the path for reproducible spin transport experiments in carbon allotropes other than graphene, such as carbon nanotubes, which have eluded researchers for more than a decade.

We acknowledge the financial support from the Spanish Ministerio de Econom\'{\i}a y Competitividad MINECO (MAT2010-18065) and the European Research Council under the European Union's Seventh Framework Programme (FP7/2007-2013) / ERC Grant agreement 308023 SPINBOUND. M.V.C acknowledges support from MINECO (RYC-2011-08319) and J.F.S. from AGAUR (Beatriu de Pin\'{o}s program).

\end{document}